# How and why silicon sensors are becoming more and more intelligent?

**Daniela Bortoletto**[a]

[a] *Physics Department, University of Oxford*
*Keble Road, Oxford, UK*
*E-mail*: Daniela.Bortoletto@physics.ox.ac.uk

ABSTRACT: This paper describes the historical evolution of silicon detectors from simple strip configurations to hybrid pixel detectors for high energy physics applications. This development has been critical to maintain the necessary physics performance in the high-rate and high-density environment of the LHC. The importance of pixel detectors and their evolution for future projects and other fields is also described.

## Contents

## 1. Introduction

Silicon is the dominant semiconductor material used in the production of position sensitive detectors for particle physics. The moderate band gap between the conduction and the valence band of 1.12 eV is large compared to the thermal energy at room temperature of 25.9 meV. Therefore cooling is necessary only in ultra-low noise applications or when required to mitigate radiation damage. The detection of minimum ionizing particles (MIP) is based on ionisation or excitation of atoms in the medium caused by the passage of charged particles. The energy required to create an electron-hole (e-h) pair is 3.6 eV yielding an ionization of about 80 e-h/μm. Thus silicon detector can be quite thin compared with gaseous detectors. The typical thickness used in high-energy physics varies between 100 and 500 μm.

In intrinsic silicon there are $\sim 10^9$ free charge carriers at room temperature but only $\sim 2 \times 10^4$ electrons are induced by a MIP traversing 300 μm. Therefore a MIP signal would be lost among the large number of free charge carriers. The operation of silicon detectors requires sensors to be fully or partially depleted of free charge carriers. This can be achieved by using p-n junctions operated in reverse bias. Furthermore position information can be achieved by finely segmenting the p-n junctions into strips or pixels using the planar technique [1].

Silicon sensors are often fabricated on n-type bulk by adding a type V material, like phosphorus (donor impurity) to silicon. Donor impurities provide an excess of electrons charge carriers. Similarly a “p-type” material can be realized by adding type III material like boron (acceptor impurities) that yields an excess of holes as majority charge carriers. Typical doping concentration used in n-bulk HEP silicon sensors are of the order of $10^{12}$ $cm^{-3}$ while the strip/pixel implant doping, which is denoted as $n^+$ or $p^+$ is of the order of $10^{14}$–$10^{16}$ $cm^{-3}$. The full depletion voltage $V_{FD}=D/2\varepsilon\mu\zeta$ depend on the thickness D, the resistivity ε, the carrier mobility μ, and the shape of the junction ζ [2]. Before irradiation detectors can be fully depleted by applying a reverse bias of about 100 V, such that only thermally-generated currents contribute to the leakage current.

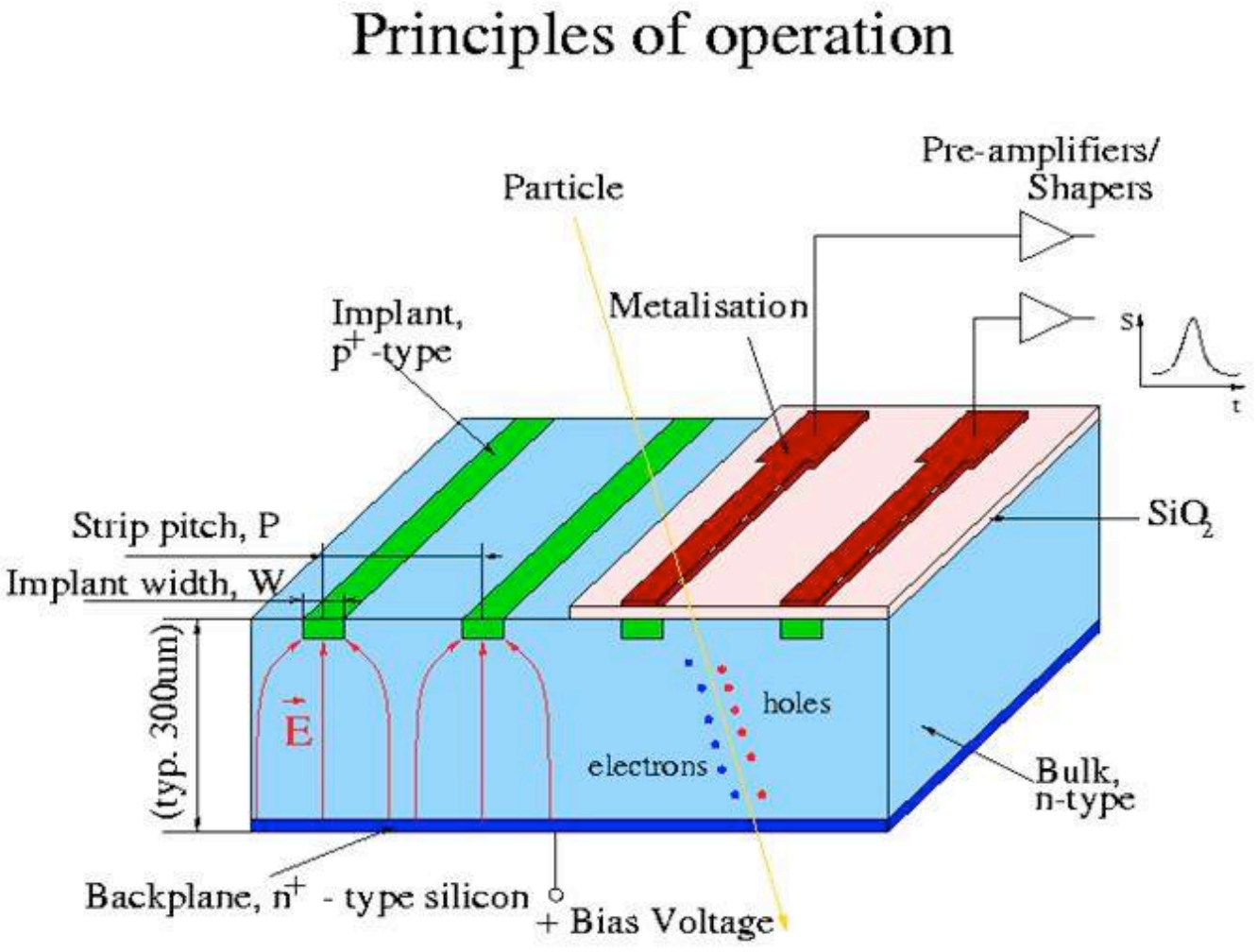

**Figure 1: Principles of operation of a silicon strip detector**

The principle of operation of a silicon strip detector is illustrated in Fig. 1. Typical configurations include $p^+$-n, $n^+$-p and $n^+$-n sensors. In a $p^+$-n configuration the generated holes drift along the electrical field, created by the bias voltage, to the $p^+$ doped strips while the electrons drift to the $n^+$ backplane. For LHC applications where radiation-damage is especially critical $n^+$-p or n+-n sensors [3] that can be operated under-depleted are used. Furthermore in $n^+$-p or $n^+$-n sensors electrons are collected which are less sensitive to trapping generated by radiation effected due to their higher mobility. The area of the detectors use in HEP are limited to the standard wafer sizes used in high-resistivity processing by industry, which has increased the wafer size from 4” to 6”. Investigation of production on larger wafers is ongoing.

## 2. Strip Sensors

The use of silicon strip detectors in HEP particle started in fixed target experiments at CERN [4] and FNAL [5] to study particles containing the charm quark. The theoretical expectations in the 1970s for the lifetime ($\tau$) of the lowest mass charm states were of the order of 0.1 ps ($c\tau$ 30 μm) with a production cross section of a few microbarns. Therefore charm events should exhibit a primary vertex and at least a secondary vertex with a separation between primary and secondary vertices of the order of $\beta\gamma c\tau$, $\beta\gamma$ being the relativistic factor of the charm state and $\tau$ its lifetime. The detectors necessary to study charm were required to have spatial resolution better than 10 μm, good particle separation, rate capability of about $10^6$Hz and low multiple scattering and photon conversion. As can be seen from Table 1, which summarizes the state of the art technologies available in 1979, there was no technology that could yield the good position resolution and low dead time necessary to study charm. The introduction of silicon detectors, which achieve excellent spatial resolution and can be operated at much higher interaction rates, was critical for studying not only charm but all heavy quarks.

Silicon strip sensors processed with planar technology were adopted for the NA11 experiment [6] at the SPS at CERN and the E706 [7] experiment at FNAL to identify and measure charm mesons properties. NA11collectect $\pi$ −Be data at 120, 175 and 200 GeV at the SPS. E706 was designed to perform a comprehensive study of direct photon production in hadron-induced collisions. A common characteristic of the early silicon detectors was the small dimension of the active areas in comparison with that of the readout components. In Fig. 2 (left) we can see that the 2.4 cm x 3.5 cm NA11 detectors are visible and they are clearly much smaller than the fan-out needed to bring the signals to the large banks of amplifier boards with discrete components.

| Chamber Type | Accuracy (rms) | Resolution Time | Dead Time |
|---|---|---|---|
| Bubble | ±75 μm | ≈1 ms | ≈1/20 s |
| Streamer | ±300 μm | ≈2 μs | ≈100 ms |
| Optical Spark | ±200 μm | ≈2 μs | ≈10 ms |
| Magnetostrictive Spark | ±500 μm | ≈2 μs | ≈10 ms |
| Proportional | >±300 μm | ≈50 ns | ≈200 ns |
| Drift | ±50-300 μm | ≈2 ns | ≈100 ns |

**Table 1: Detectors performance from the 1978 Particle Data Group**

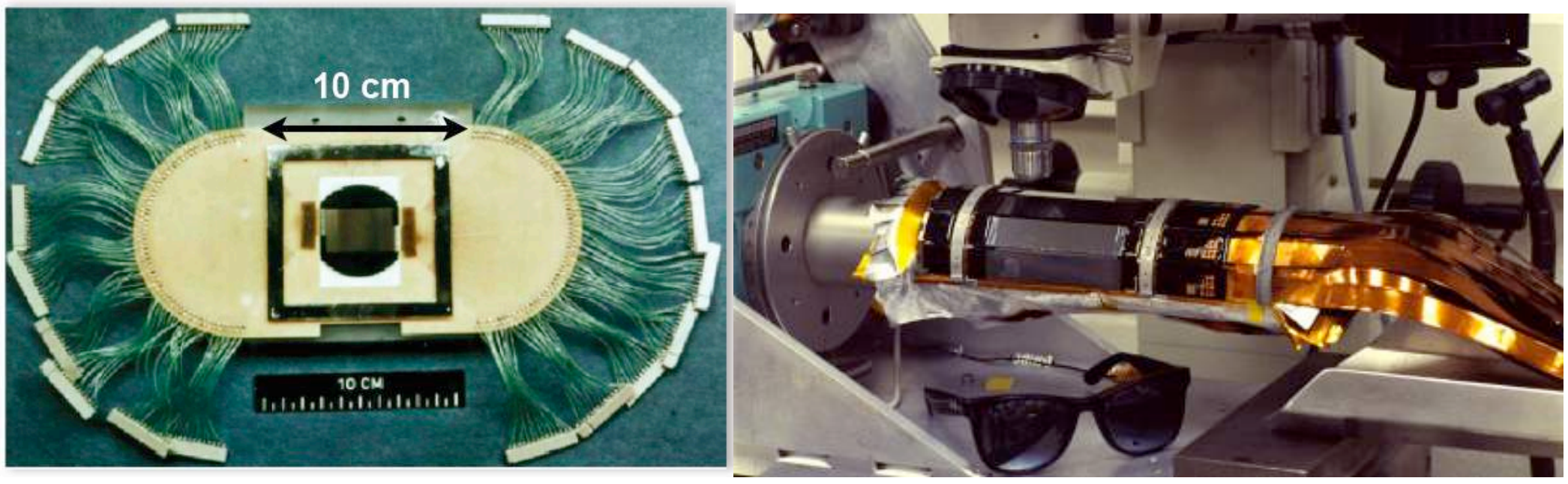

**Figure 2: Photographs of a mounted NA11 Detector The sensor seen in the centre is 24×36 mm$^2$ in size (left) and of the MARK II vertex detector (right).**

In order to insert silicon detectors into collider experiments miniaturization of the electronic was necessary. This was achieved with the development of custom VLSI readout electronics that could be coupled directly to the detectors [8]. One of the first examples of this advancement was the silicon detector developed for the Mark II [9] at the SLAC [10] Linear Collider (SLC), which is shown in Fig. 2 (Right). The goal of this 3-layer silicon strip detector was to provide a few high precision point measurements as close to the beam-beam interaction point as possible. These precision measurements would then be complement the angle and momentum measurements in a drift chamber. With an expected position resolution of about 5μm per point, the addition of a silicon vertex detector significantly improved the measurement of the impact parameter (Fig. 3, left) of high momentum charged tracks relative to the beam-beam interaction point. In addition, the fine granularity of the silicon strip detectors provided the ability to resolve tracks that are separated by as little as 5mrad reducing errors in track pattern recognition. As a result of this advancement MARK II provided one of the first measurement of the B meson lifetime by partially reconstructing B mesons decaying into $D^{*-}$ mesons plus charged leptons. The measurements of the impact parameter of the lepton showed that it was produced a small fraction of a millimetre away from the collision point. This offset depends on the distance travelled by the heavy particle before it decayed, and hence on the lifetime of the original particle.

In the 1990s vertex detectors enabled the study of heavy flavor physics both in fixed target and colliding beam experiments. All four LEP experiments, ALEPH, DELPHI, OPAL, and L3 had vertex detectors containing silicon strip detectors [11]. These vertex detectors allowed tagging the presence of heavy flavour (charmed and beauty) hadrons in the decay of the Z resonance produced at LEP. At the SLAC SLD a detectors containing charged coupled devices allowed similar studies as shown in Fig. 3 (right). The study of heavy quarks and their decays was continued with the construction of B-factories at KEK [12] and SLAC. Asymmetric b-factories use colliding beams of different energies to produce $\Upsilon(4S)$ mesons with a boost in one direction. With a mass of 10.58 GeV the $\Upsilon(4S)$ particle has just enough energy to decay into two B mesons. Because of the boost, the B mesons in an asymmetric B factory decay at measurably different points along that direction, allowing an accurate determination of B mixing and CP violation in the B-system. The presence of many low-energy particles in B decays requires that silicon detectors at B factories minimize the amount of material in the active volume.

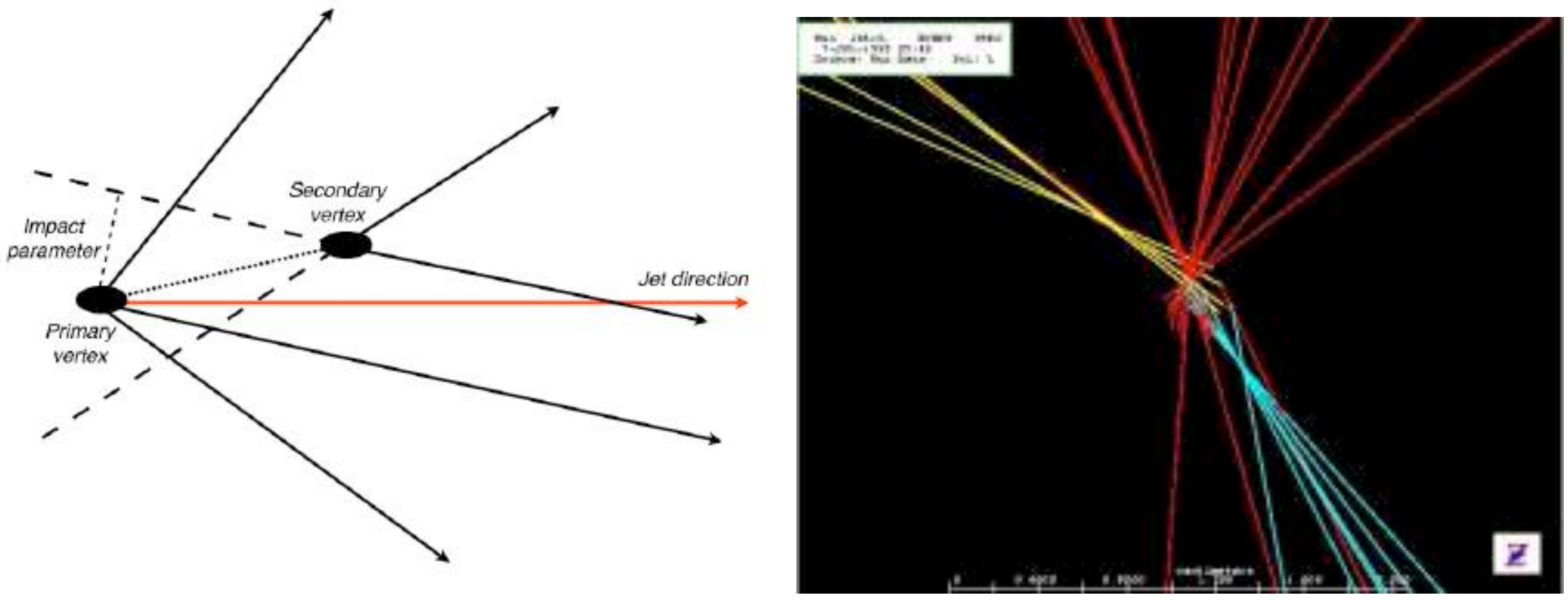

**Figure 3: Finding heavy quarks by measuring the impact parameter or reconstructing secondary vertices (left) and event from the 300 Megapixel SLD vertex detector showing a Z decay into a gluon jet (red) and two jets containing a beauty and a anti-beauty quark (blue, yellow). The heavy quark jets show detached vertices relative to the Z decay vertex.**

One of the most important developments of silicon detectors was their deployment in hadron colliders, which required the development of radiation-hard electronics and special attention to radiation damage effects. SVX, the first silicon detector at a hadron collider, was installed in CDF for the start of Run 1A of the Tevatron in 1992 [13]. SVX could measure tracks with a precision of tens of microns, good enough to identify B decays, for which have a typical decay length $c\tau=450\mu m$ times the boost due to the motion of the B meson. By 1994 CDF found six b-tagged events consistent with top decays into a W boson and b quark over a background of 2.3±0.3. The evidence for the top quark became a discovery in 1995 when both CDF and D0 announced the discovery of the top quark. Fig. 4 shows a golden $t\bar{t}$ leading to two b-tagged jets. The SVX was not radiation hard enough and it was replaced first by the SVX' and then by the SVX II and ISL detectors. The SVX circuit [14] was the first custom CMOS IC including sparse-data scanning technology, in which "hit" channels (those with genuine signals) were identified and processed. Sparse scanning saved about two orders of magnitude of time and suppressed useless data. This was a revolutionary advance that has become ubiquitous in the field.

Starting with CDF [15] and then D0 [16] silicon detector were used not only for vertexing but also for precision tracking including momentum determination in the magnetic field. Because of this new requirement the size of silicon detectors has been increasing, leading to more complexity for power delivery and cooling. The strip silicon trackers currently inserted in ATLAS[17] and CMS[18] are 200 and 61 $m^2$ respectively. These large-scale systems have been made possible by a factor of about 40 decrease in the price per detector area. This was made possible by increasing the wafer area from 4” to 6” wafers and by a reduction in the cost of wafer processing.

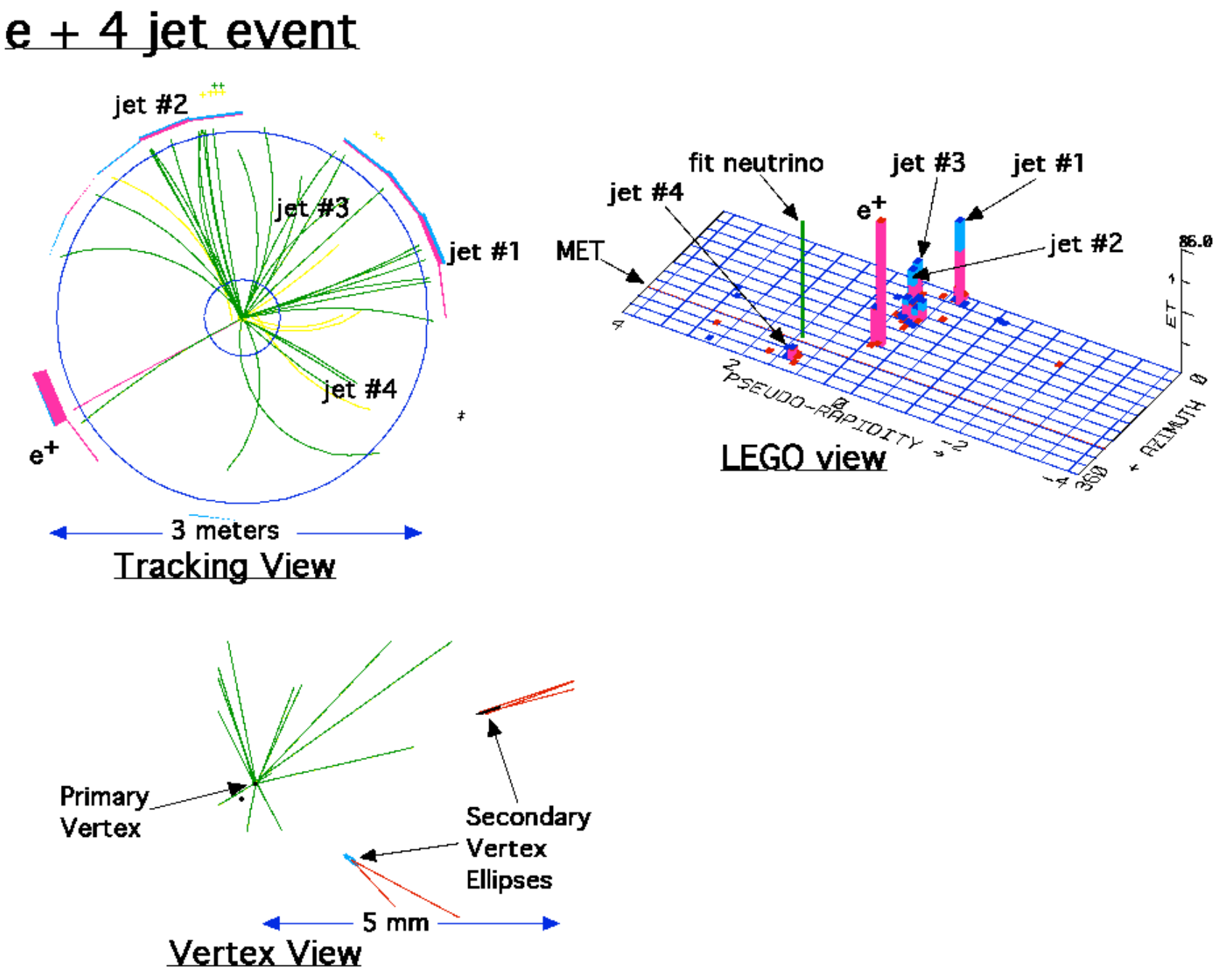

**Figure 4: A $t\bar{t}$ event with the top quark decaying into a W and a b-quark. One the W boson decays leptonically into a lepton and missing energy. The second W decays into two jets. The event has one lepton, four jets (two tagged b jets) and missing energy.**

Nonetheless the continue increase in luminosity at hadron colliders and the need of low occupancy for track reconstruction has required the development of pixel detectors for the innermost layer of CMS and ATLAS. Pixel detectors are also necessary for precision measurement at e+e- colliders.

## 3. Pixel Sensors

The development of pixel detectors has not only been critical for tracking in the extreme environment of the LHC but it has also lead to an array of applications in other fields. The hybrid pixel technology, which was initially developed for ionizing particle trackers was modified for X-ray imaging. Hybrid pixel X-ray cameras are providing significant advantages with respect to CCD cameras such as excellent signal-to-noise ratio, very high dynamic range, low-energy X-ray suppression, short readout time, and high frame rates. In addition they provide modular detectors enabling multi-module detectors with large active area. Hybrid pixel detectors are currently routinely used in cameras for material sciences (crystallography), non-destructive control, biomedical imaging and clinical imaging.

### 3.1 Hybrid pixel detectors

In the LHC environment when the number of tracks becomes very large and close to the beam interaction point, the probability that two particles hit same strip sensor (occupancy) is high. Furthermore even using a double-sided strip detector it is not possible to determine which pair of strips corresponds to a particle as shown in Fig. 5 (Left). This was already a problem for the LEP experiments particularly in the forward direction. This problem becomes more critical for proton collisions at the LHC, as shown in Fig. 5 (left), which displays a Higgs event in ATLAS from run1 of the LHC. Another problem at the LHC is the degradation of the sensor due to the high radiation doses. Pixel detectors are more robust in this environment since they offer high signal to noise ratio due to the small capacitance at the input of the amplifier, which is not the case for long strips.

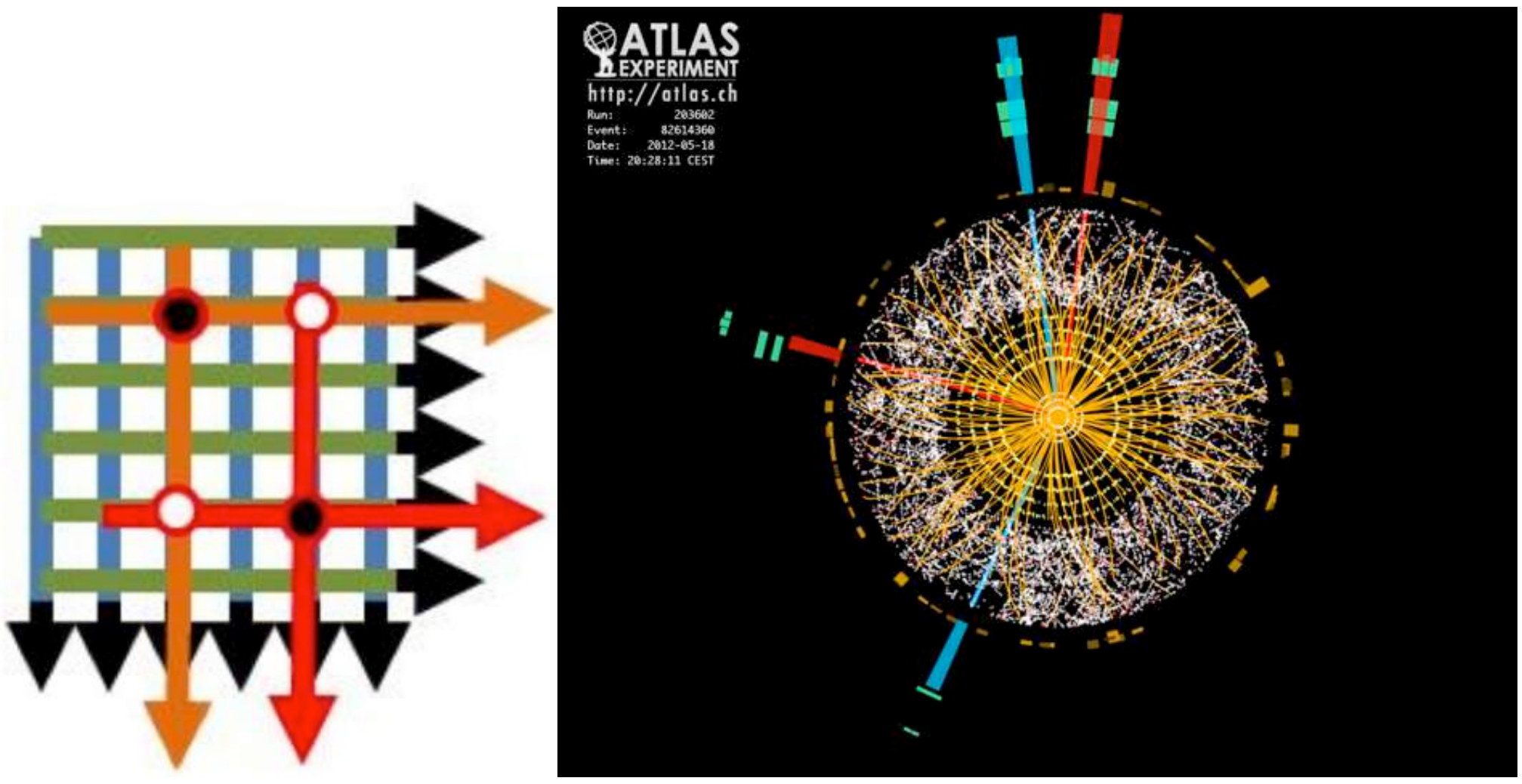

**Figure 5: A sketch of ambiguities and ghost hits in double-sided strip detectors (left). Higgs event in ATLAS**

The basic building block of a hybrid pixel detector is the combination of a monolithic matrix of reverse biased silicon diodes connected to a number of large area CMOS readout Application Specific Integrated Circuits (ASICs) as shown in Fig. 6 (left). The first paper presenting the feasibility of such a detector was published in 1988 [19], Consequently significant R&D took place within the RD19 framework created in 1990 [20] to support the development of pixel detectors for the LHC. In 1995 the first fixed target pixel detector was installed in the WA97 (CERN, Omega facility) and, almost at the same time, the first collider pixel detector was installed in the DELPHI experiment at LEP [21].

The main challenge of hybrid detectors is that the electronics must cover the same area of a single pixel, which is about a hundred micrometer square. The contact between the sensor pixel and the input of the electronics was achieved by using the bump-bonding technique, which allows the placement of one solder ball per pixel input pad. Then these solder balls are fused once the ASIC is positioned on the sensor. Because of the high density of transistors (several millions of transistors per square centimeter), the integrated circuit cannot be large, whereas the sensor can be. Therefore several ASICs can be bumped to one sensor. A photo of several forward CMS pixel modules is shown in fig. 6 (right).

Pixel detectors are critical to the success of the LHC physics program. To obtain the high collision rates necessary to generate the rare events of interest the two rings of the LHC are filled with about 1380 bunches containing $1.7\times10^{11}$ protons per bunch and colliding 20 million times per second. Every time two bunches of protons cross, they generated not one collision, but on average about 37 collisions per crossing at the highest rates obtained during the LHC first run (Run1). This effect is called pile-up. Pile-up obscures the reconstruction of the events of interest. With cell size of 50 $\mu$m$\times$400$\mu$m (ATLAS) and 100 $\mu$m$\times$150$\mu$m (CMS), pixel detectors attain the position resolution necessary for distinguishing detached and multiple vertices in this complex environment. In addition, because their granularity, tracks are unlikely to traverse the same pixel and therefore pixel detectors are essential to reconstruct the trajectories of tracks emerging from these complex collisions and measure their momentum.

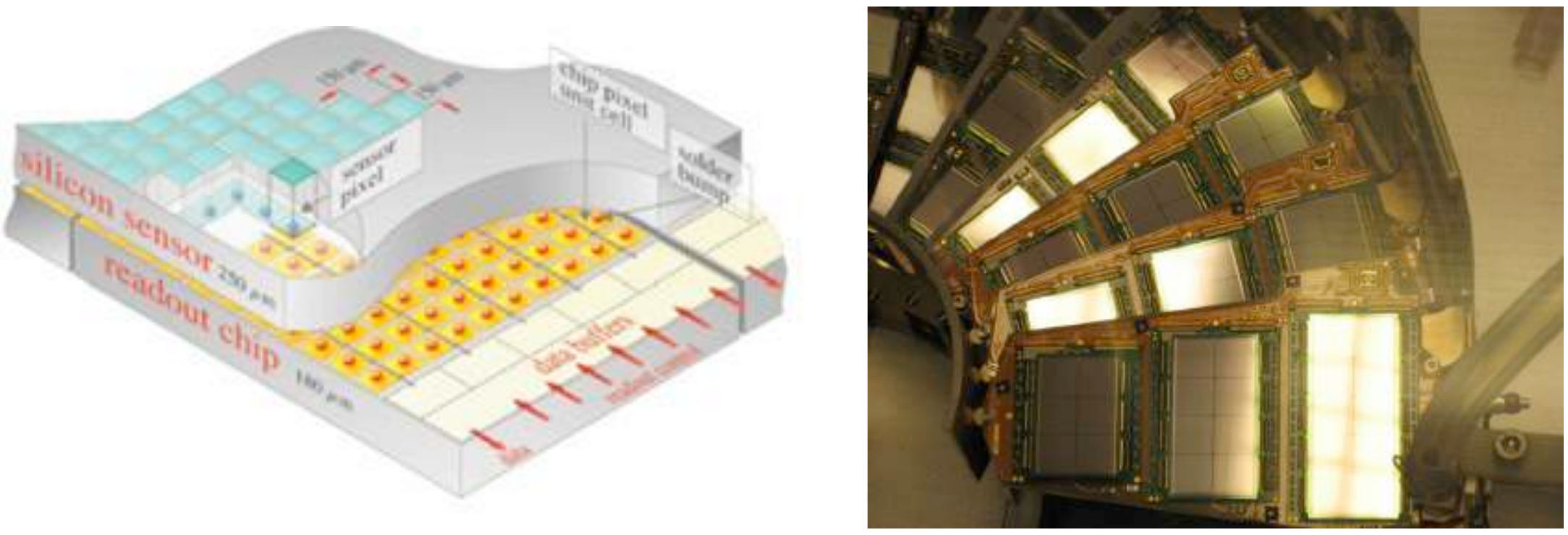

**Figure 6: A sketch of a hybrid pixel detector (left). A photo of the CMS forward pixel detector blades containing modules with different number of ASICs (right).**

The LHC pixel detectors were developed and built from 1998 and 2008. The ATLAS [22] and CMS [23] pixel detectors contain 80 and 66 million pixels. The ALICE experiment for heavy ion collisions has a 10 million pixels [24] detector. Silicon sensors bump-bonded to ASICs and mounted on high density interconnects are shown in Fig. 6 (right). The sensor, which is made of high resistivity material, is operated in full depletion allowing the charge generated by charged particles to be collected within the 50 ns bunch crossing of the LHC. The readout circuit compensates radiation induced detector leakage current and integrates the charge pulse associated with the particle traversing the silicon. In addition it filters the resulting signal to minimize the electronic noise and it determines if the signal is above a set threshold. If the hit passes the requirements its arrival time is associated to a given bunch crossing. Some ASICs also record the amplitude of the collected charge. A typical pixel readout channel comprises of a few hundred transistors. The ASICs used during run1 of the LHC are designed in a 250 nm CMOS process applying design techniques that minimize the radiation-induced degradation of performance throughout the expected lifetime [25].

The LHC pixel detectors have performed extremely well in the challenging LHC environment and have been essential to extract the LHC physics. For example the resolution of the CMS pixel detector measured in the middle barrel layer by comparing the predicted position with the measured hit for tracks with $p_T > 12$ GeV, which are not affect by multiple scattering in the other layers, is around 9.4 $\mu$m. The primary-vertex reconstruction determines the location of all proton-proton interaction vertices in an event. The resolution in jet-enriched samples approaches 10 $\mu$m in x and 12 $\mu$m in z for primary vertices using at least 50 tracks. A similar performance was achieved by the ATLAS pixel detector. The efficiency of tracks having associated hits in the different pixel detector layers was about 99% and the typical occupancies for physics data have been in the order of $10^{-4}$ for both for ATLAS and CMS.

### 3.2 Hybrid pixel for X-ray detection

The idea of counting single X-ray photons was already being explored in the late 90's [26]. Hybrid pixels are ideally suited for this applications offering lower noise, speed, efficiency and the unique possibility of energy selection. Nonetheless these applications require a change in the ASIC readout architecture. Several ASICs, MEDIPIX [27], XPAD[28], PILATUS [29] and PIXIE[30], have been developed often starting from circuits for particle physics experiments. For X-ray imaging instead of selecting individual frames for readout, all hits above threshold in a given pixel are counted and then the entire accumulated image is sent off chip. X-rays below ~10keV are detectable in standard high resistivity silicon. However, in many medical applications 20keV-140keV X-rays are used to form an image. In this case sensors using GaAs instead of silicon are used. PILATUS, a spin-off of the PSI chip developed for the CMS hybrid pixel detector, is available from Dectris [31] and it has become a standard for systems used in synchrotron radiation applications. Following these developments several prototype ASICs have also been developed by companies such as Siemens, Philips, and Toshiba for application for Computed Tomography (CT) scanners.

## 4. Fully depleted Thick CCD

An extended response in the red region of the electromagnetic spectrum is extremely important for charge couple devices (CCDS) used in astrophysics. This requirement and ideas from particle physics detectors has lead to the development of fully depleted charge-coupled devices (CCDs) [32]. Standard scientific CCDs are thinned and back illuminated to achieve high quantum efficiency (QE). Thinning is required, in this case, because the low-resistivity silicon limits the depth of the depletion region and degrades the red response.

.

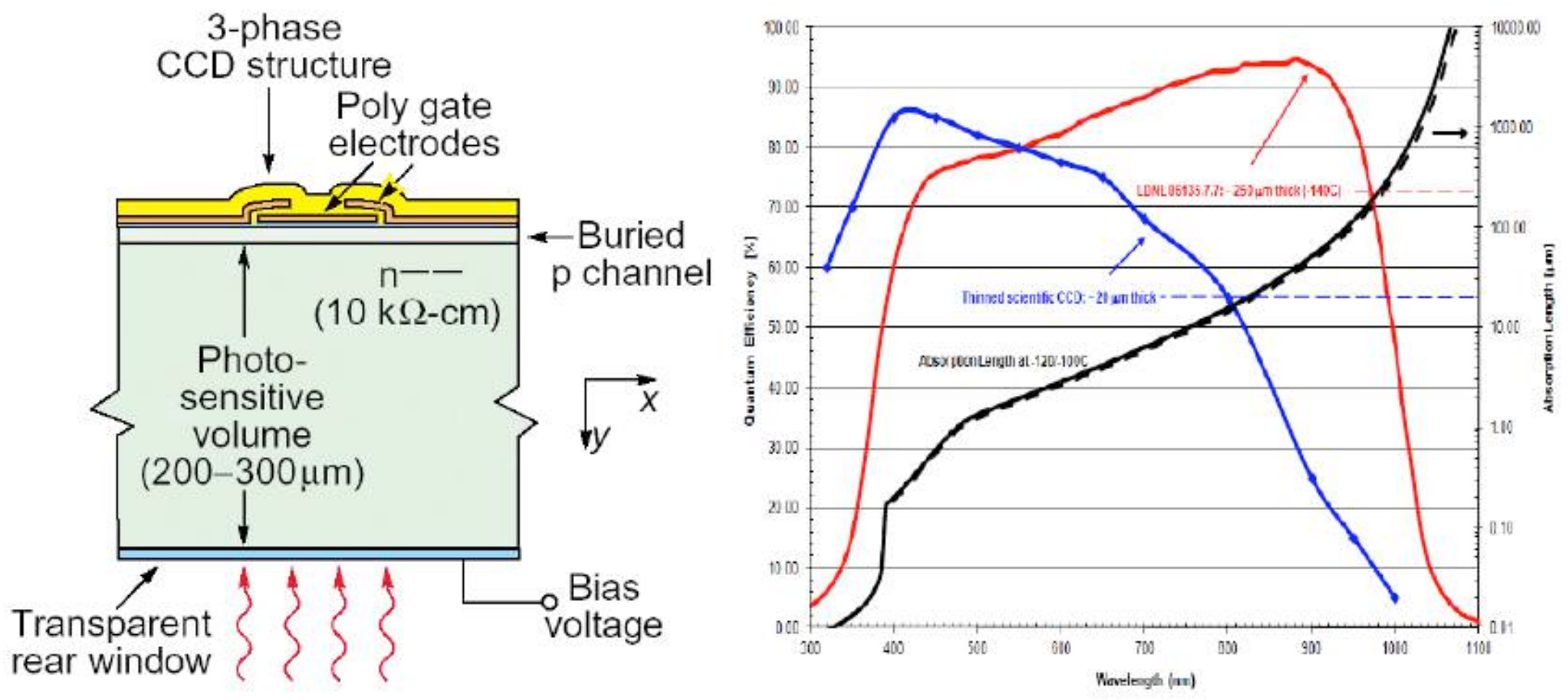

**Figure 7; Sketch of thick CCD (left) and Quantum efficiency of fully depleted thick CCDs with respect to standard scientific CCDs**

Thick CCD are fabricated on high-resistivity n-type silicon substrates (>5000 Ω-cm) which can be biased to have a 200–300 µ m thick fully depleted sensitive region. Fig. 7 (left) shows a cross-sectional diagram of the fully depleted, back-illuminated CCD. The thickness of the CCD results in improved near-infrared sensitivity when compared to conventional thinned CCD's as shown in Fig. 7 (right), which shows the measured quantum efficiency (QE) versus wavelength for a fully depleted, back-illuminated CCD operated at – $130^0$ C. The results demonstrate an improvement of a factor of 5 around 900 nm, which is very important to study dark energy.

## 5. New Directions

The Phase II ATLAS and CMS detectors are being designed to cope with the very challenging beam conditions of the LHC after 2022. The High Luminosity LHC (HL-LHC) will operate at a higher center-of-mass energy of 14 TeV and collision rates about a factor of 5-10 higher than the one for which the current detector were built. Up to now about 20 $fb^{-1}$ of data have been collected. By the end of the HL-LHC program about 3000 $fb^{-1}$, a data sample 150 times larger than the one of run1, will be delivered to the experiment. The high instantaneous luminosity will come at the price of extremely high pileup. Up to 140 overlapping events for a bunch-crossing interval of 25 ns are foreseen. In order to operate in such environment the ATLAS and CMS tracker and pixel detectors must be replaced. Especially in the layers closer to the interaction region, the next generation detectors must achieve high rate capability using integrated circuits and materials with increased tolerance to radiation dose and single event upsets (SEU). The latter are changes of state in a micro-electronic device caused by ions or radiation striking a sensitive node. For pixel vertex detectors the HL-LHC requirements translate to a need to develop sensors capable of surviving doses of about $2\times10^{16}$ $n_{eq}/cm^2$ and to handle hit rates of about $1GHz/cm^2$.

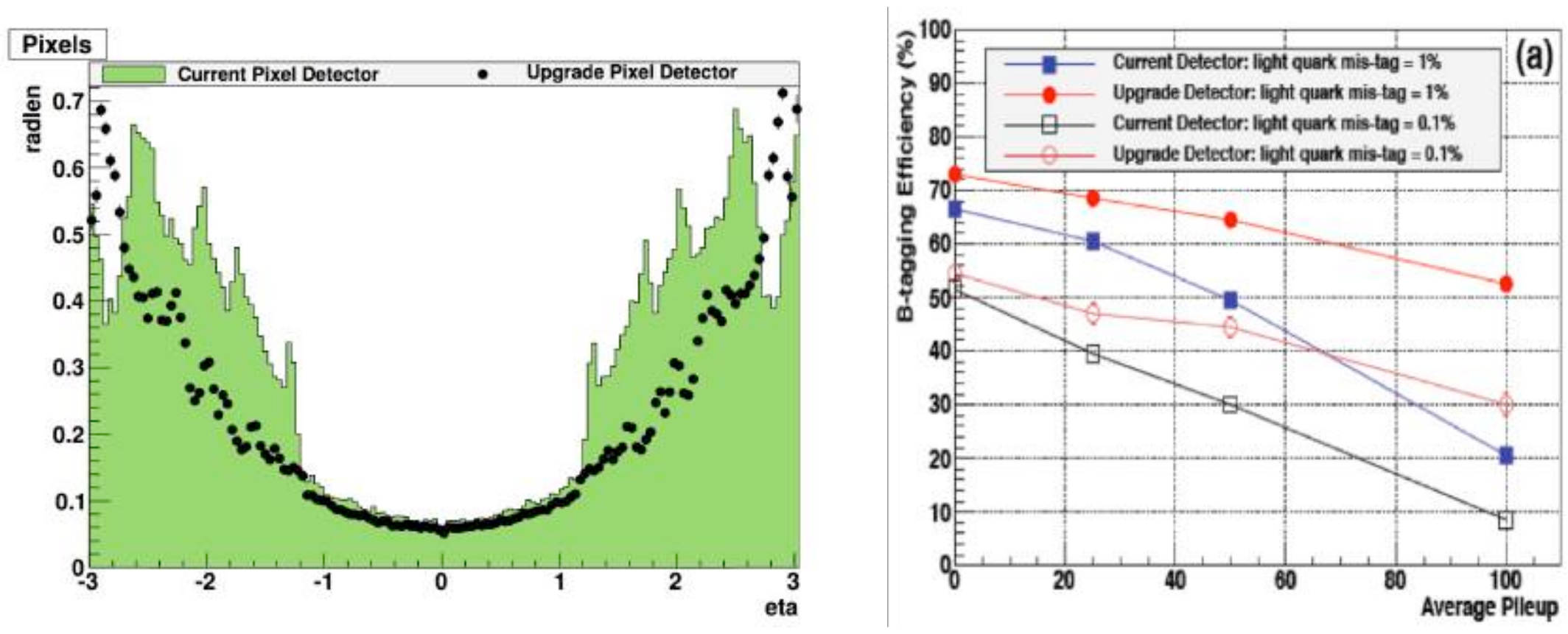

**Figure 8: Material distribution for the CMS Phase 1 detector as a function of η (left) and b-tagging efficiency versus pile-up for the upgrade and the current detector for different probabilities of a light jet to fake a b-jet.**

A new R&D effort (RD53) [33] has been organized to develop the ASICs for the Phase 2 detectors. In the case of the upgrades for ATLAS [34] and CMS [35] the plan is to use 65 nm CMOS technology and modify the readout ASIC architectures to be able to deal with the higher particle fluxes expected. Development is taking place to render sensors more radiation hard by making them thinner [36] or by using the 3D architecture [37]. There is also a lot of effort in the optimization of the full detector performance [38]. This is already evident in the ATLAS IBL[39] and development of the CMS pixel detector for an intermediate upgrade denoted as Phase I [40]. The Phase 1 pixel detector will be installed in CMS before or during the second long shutdown of the LHC, which is currently planned for 2018. A lot of development for the Phase 1 CMS pixel detectors was focused on advanced mechanics to reduce the material and improve the tracking and b-tagging performance as shown in fig 8.

The LHCb upgrade will require a readout architecture that allows all hit time stamps and amplitudes to be sent off chip independently of an external trigger. The Alice experiment focuses on heavy ion physics that requires interaction rates and total accumulated radiation doses much smaller than those for Atlas and CMS. ALICE will use ultra-thin monolithic [41] devices (Fig. 9 (left)) where sensors and readout ASICs are incorporated onto the same substrate for its upgrade. The upgraded ALICE Inner Tracking system (ITS) shown in Fig. 9 (right) will improve the track position resolution at the primary vertex by a factor of 3 or even larger with respect to the present detector. Monolithic pixel sensors and special carbon fibre support structures will reduce the material and improve tracking performance. Monolithic pixel use special semiconductor processes developed for CMOS sensors.

High-Voltage and High-resistivity CMOS processes [42] are under consideration also for the phase 2 ATLAS strip and pixel tracker upgrade. The development of monolithic pixel and HV/HR CMOS for particle physics is especially interesting for future trackers. The reduction of mass is critical for the high precision vertex detectors for future $e^+e^-$ linear collider. The cost reduction that could be achieved by building monolithic pixel detector could be essential for the trackers at future 100 TeV hadron colliders.

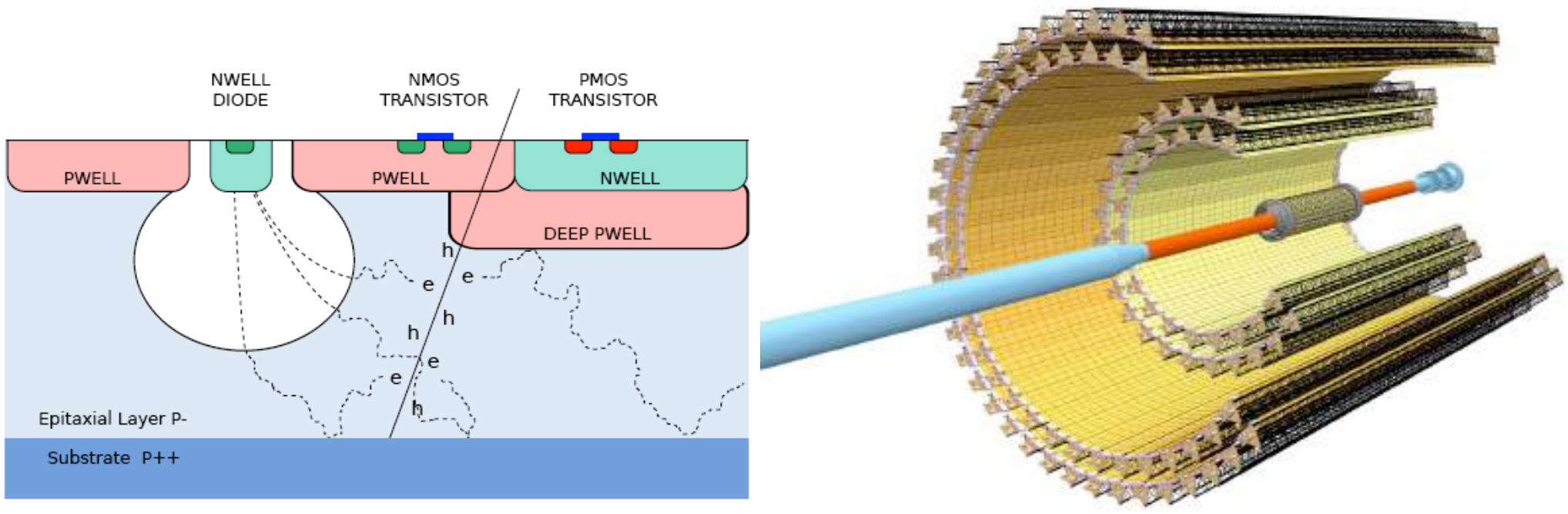

**Figure 9: A sketch of a monolithic pixel detector (left ) and the layout of the ALICE tracker upgrade (right)**

## 6. Acknowledgments

The author would like to thank the INFIERI network for organizing a great workshop in Paris. She also acknowledges several review papers on silicon sensors from P. Delpierre [43], H. Sadrozinski [44], M. Campbell [45], and F. Hartman [2].

## References

[1] J. Kemmer, *Fabrication of a low-noise silicon radiation detector by the planar process*, Nucl. Instrum. Meth. A169, (1980) 499.

[2] F. Hartman, *Evolution of Silicon Sensor Technology in Particle Physics*, Springer Science & Business Media, 2009.

[3] G. Bolla et al., *Sensor development for the CMS pixel detector*, Nucl. Instrum. Meth. A485, 89, 2002; T. Rohe, et al., Nucl. Instr. and Meth. A 460 (2001) 55; Nucl. Instr. and Meth. A 456 (2001) 217.

[4] CERN: http://home.web.cern.ch/.

[5] FNAL: http://www.fnal.gov/.

[6] J. Kemmer, E. Belau, R. Klanner, G. Lutz, B. Hyams, *Development of 10um resolution silicon counters for charm signature observation with the ACCMOR spectrometer*, in Batavia 1981, Proceedings Silicon Detectors For High Energy Physics*, (1981) 195.

[7] E. Engels, et al. (E706 Collaboration), *A silicon microstrip vertex detector for direct photon physics, Nucl. Instrum. Meth.* A253, (1987) 523.

[8] J. T. Walker, S. Parker, B. Hyams, S. L. Shapiro, *Development of high density readout for silicon strip detectors*, .Nucl. Instrum. Meth. A226, (1984) 200.

[9] C. Adolphsen, R. Jacobsen, V. Luth, G. Gratta, L. Labarga, A. Litke, A. Schwarz, M. Turala, C. Zaccardelli, A. Breakstone, C.J. Kenney, S.I. Parker, B.A. Barnett, P. Dauncey, D. Drewer, J.A.J. Matthews, *The Mark-II silicon strip vertex detector*, Nucl. Instrum. Meth. A313, (1992) 63.

[10] SLAC National Accelerator Laboratory: https://www6.slac.stanford.edu/

[11] A.S. Schwarz, *Silicon strip vertex detectors at LEP*, Nucl. Instrum. Meth. A342, (1994) 218.

[12] KEK High Energy Accelerator Research Organization: http://www.kek.jp/en/.

[13] S. Tkaczyk et al., *The CDF Silicon Vertex Detector*, Nucl. Instrum. Meth. A 342, (1994) 240.

[14] S. A. Kleinfelder et al., *A flexible 128 channel silicon strip detector instrumentation integrated circuit with sparse data readout*, IEEE Transactions on Nuclear Science, 03/1988.

[15] T. Altonenen et al., *Operational experience, improvements, and performance of the CDF Run II silicon vertex detector*, Nucl. Instrum. Meth. A 729, (2013) 143.

[16] J. Fast, *The D0 silicon detector for run IIb at the Tevatron*, Nucl. Phys. B, (2003)352.

[17] G**.** Aad et al, *The ATLAS Experiment at the CERN Large Hadron Collider*, JINST 3 (2008) S08003.

[18] S. Chatrchyan et al, *The CMS experiment at the CERN LHC*, JINST **3,** (2008) S08004.

[19] E.H.M. Heijne, P. Jarron, A. Olsen and N. Redaelli, *The silicon micropattern detector: a dream?* , Nucl. Instrum. Meth. A 273, (1988) 615.

[20] CERN RD19 collaboration, *Development of silicon micropattern pixel detectors*, Nucl. Instrum. Meth. A 348, (1994) 399.

[21] K.H. Becks et al., *The DELPHI pixels*, Nucl. Instrum. Meth. A 386, (1997) 11.

[22] V. Vrba, Nucl. Instr. and Meth. A 465, (2001) 27.

[23] D. Bortoletto, *The CMS pixel system*, Nucl. Instr. and Meth. A 579, (2007) 669.

[24] F. Meddi, Nucl. Instr. and Meth. A 465, (2001) 40.

[25] G. Anelli, M.Campbell, M. Delmastro et al, IEEE Trans. Nucl. Sci. Vol 46 pp 1690-1696 (1999).

[26] M. Campbell, E.H.M.Heijne, G. Meddeler, E. Pernigotti, S. Snoeys, *Readout chip for a 64x64 pixel matrix with 15-bit single photon counting*, IEEE Nucl. Sci. Symp. (1998);189 P. Fischer, J. Hausmann, M. Overdick et al., *A counting pixel chip and sensor system for X-ray imaging*, Nucl. Instr. and Meth. A405 (1998) 53.

[27] R. Ballabriga, G Blaj, M Campbell, M Fiederle et al., *The Medipix3RX: a high resolution, zero dead-time pixel detector readout chip allowing spectroscopic imaging*, J. Instr. 2011 1748-0221 6 C01052.

[28] L. Blanquart et al., *XPAD, a new read-out pixel chip for X-ray counting*, Nuclear Science Symposium Conference Record, 2000 IEEE, vol.2, no.pp.9/92-9/97 vol.2, 2000.

[29] P. Kraft, A. Bergamaschi, C. Bronnimann, R. Dinapoli et al., *Characterization and Calibration of PILATUS Detectors*, IEEE Trans. Nucl. Sci. Vol. 56 (2009) 758.

[30] PIXIE: http://www.pixirad.com/

[31] Dectris: https://www.dectris.com/

[32] S. Holland et al., *Fully depleted back-illuminated Charge-coupled devices fabricated on high-resistivity silicon*, IEEE transaction on Electron Devices 50, (2003) 225.

[33] J. Christiansen, M. Garcia-Sciveres, *RD Collaboration proposal: Development of pixel readout integrated circuits for extreme rate and radiation*, CERN-LHCC-2013-008 LHCC-P-006 (2013).

[34] ATLAS Collaboration, *ATLAS Letter of Intent Phase-II Upgrade*, CERN-LHCC-2012-022, LHCC-I-023 (2012).

[35] CMS Collaboration, *Technical proposal for the upgrade of the CMS detector through 2020*, CERN-LHCC-2011-006 CMS-UG-TP-1 LHCC-P-004 (2011)

[36] S. Terzo et al. *Heavily Irradiated N-in-p Thin Planar Pixel Sensors with and without Active Edges*, (2014) arXiv:1401.2887 [physics.ins-det]

[37] G-F. Dalla Betta, *et al. Recent developments and future prospectives in 3 D radiation sensors*, Journal of Instrumentation, 7 (2012), C10006

[38] C. Da Via, *3D silicon sensors: Design, large area production and quality assurance for the ATLAS IBL pixel detector upgrade*, Nucl. Instr. and Meth. A 694 (2012) 321.

[39] D. Bortoletto, *The importance of silicon detectors for the Higgs boson discovery and the study of its properties*, Mod. Phys. Lett. A **29**, (2014) 1430041.

[40] CMS Collaboration, *CMS Technical Design Report for the Pixel Detector Upgrade***,** CERN-LHCC-2012-016; CMS-TDR-11 (2012).

[41] ALICE Collaboration, *Technical Design Report for the Upgrade of the ALICE Inner Tracking System***,** CERN-LHCC-2013-024; ALICE-TDR-017 (2013)

[42] I. Peric, *Active pixel sensors in high-voltage CMOS technologies for ATLA*S, (2012) JINST7 C08002; I. Peric and C. Takacs, *Large monolithic particle pixel detectors in high voltage CMOS technology*. Nucl. Instrum. Methods A 624 (2010), 504. R. Turchetta et al., *A monolithic active pixel sensor for charged particle tracking and imaging using standard VLSI CMOS technology,* Nucl. Instrum. Methods A 458 (2001), 677.

[43] P. Delpierre, *A history of hybrid pixel detectors, from high energy physics to medical imaging,* (2014) *JINST* **9** C05059.

[44] H. Sadrozinski, *Applications of Silicon Detectors*, IEEE Trans. Nucl. Sci., VOL. 48, NO. 4 (2001).

[45] M. Campbell, *Detecting elementary particles using hybrid pixel detectors at the LHC and beyond*, IEEE Nucl. Sci. Symp. Conf. Rec. (2014).